\RequirePackage{fixltx2e}
\documentclass[a4paper]{isp}
\usepackage[american]{babel}
\usepackage{graphicx}
\usepackage{amsmath,amsthm,mathtools}
\usepackage{url}
\usepackage[capitalise,nameinlink]{cleveref}
\usepackage{xcolor}
\def\aap{A\&A\,  }

\def\apj{ApJ\,  }
\def\apjl{ApJ Letters,  }
\def\apss{Astrophysics and Space Science  }
\def\azh{Astronomicheskii Zhurnal\,}

\begin{document}
\pdfgentounicode=1
\title
{
Energy Conservation in the thin  layer approximation:
I. The spherical classic case for supernovae remnants 
}
\author{Lorenzo Zaninetti}
\institute{
Physics Department,
 via P. Giuria 1, I-10125 Turin, Italy \\
 \email{zaninetti@ph.unito.it}
}

\maketitle

\begin {abstract}
The thin layer approximation applied to the expansion of 
a supernova remnant assumes that all the swept mass 
resides in a thin shell.
The law of motion in the thin layer approximation
is therefore found using the conservation 
of momentum.
Here we instead introduce the conservation of energy in the framework
of the thin layer approximation.
The first case to be analysed is that of an interstellar medium
with constant density
and the second case is that of 7 profiles of decreasing 
density with respect to the centre of the explosion.
The analytical
and numerical results are applied to 4 
supernova remnants:
Tycho, 
Cas A, 
Cygnus loop,
and  SN~1006.
The back reaction due to the radiative losses for 
the law of motion is evaluated in the case of constant
density of the interstellar medium.
\end{abstract}

\section{Introduction}

The thin layer approximation  assumes that the mass ejected 
in the explosion of a supernova (SN) resides  in a thin layer.
This  approximation is usually applied in the late  stage  
of the explosion in order to explain the supernova remnant (SNR),
see \cite{Bisnovatyj-Kogan1982,Tenorio-Tagle1987,MacLow1988}.
The physical quantity which  is conserved in the previous  approaches
is the momentum, equal to the swept mass 
multiplied by the velocity at a given radius of expansion $r_0$ 
equated  to these quantities at a radius  $r$.
Some  natural questions therefore arise:  
\begin{itemize}
\item 
Can we model 
the expansion of an SNR when the energy is conserved
rather than the momentum? 
\item
Can we model the energy conservation when 
the density of the interstellar medium (ISM) decreases
with the distance from the point of the explosion?
\end{itemize}
In order to answer the above  questions,
Section;  \ref{section_momentum} reviews the 
standard laws of conservation,
Section  \ref{section_energy} 
introduces the conservation of energy
and
Section  \ref{section_astro} 
applies the derived equations of motion
to 4 SNRs. 

\section{Laws of  conservation}
\label{section_momentum}
We summarise  four laws of conservation useful to model 
some  astrophysical phenomena in which 
the temperature and the pressure are absent.
The {\it first} law  is 
the conservation of  momentum in
spherical coordinates
in the framework of the thin
layer approximation.
The Newton's second law for an expanding sphere 
in the framework of the thin shell approximation 
along a solid angle $\Delta\Omega$ is  
\begin{equation}
\frac{d}{dt} \bigl( \frac{1}{3} r^3  \rho v \big ) = r^2 P
\quad ,
\end{equation}
where $r$ is the advancing radius, $\rho$ is the density
assumed to be constant, $v$ the velocity and $P$ the internal pressure,
see formula (10.27) in \cite{McCray1987}.
Let us assume $P=0$ (cold model)  and the above equation in two different points
of expansion becomes
\begin{equation}
M_0(r_0) \,v_0 = M(r)\,v
\quad ,
\end{equation}
where $M_0(r_0)$ and $M(r)$ are the swept masses at $r_0$ and $r$,
while $v_0$ and $v$ are the velocities of the thin layer at $r_0$ and $r$.
This {\it first law} has been widely used to model the SNRs,
see \cite{Kompaneets1960,Bisnovatyi-Kogan1989,Dyson1997,Bisnovatyi-Kogan1998,Padmanabhan_II_2001,Chen2003}.
This conservation law can be expressed as a differential equation
of the first order by inserting $v=\frac{dr}{dt}$:
\begin{equation}
M_0(r_0)\, v_0=M(r)\, \frac{dr}{dt}
\quad .
\end{equation}
In the case where the ISM has constant density,  
the analytical solution 
for the trajectory is 
\begin{equation} 
r(t;t_0,r_0,v_0) =
\sqrt [4]{4\,{r_{{0}}}^{3}v_{{0}} \left( t-t_{{0}} \right) +{r_{{0}}}^
{4}}
\label{rtconstant}
\quad , 
\end {equation}
and the velocity is 
\begin{equation} 
v(t;t_0,r_0,v_0) =
\frac
{
{r_{{0}}}^{3}v_{{0}}
}
{
 \left( 4\,{r_{{0}}}^{3}v_{{0}} \left( t-t_{{0}} \right) +{r_{{0}}}^{4
} \right) ^{3/4}
}
\label{vtconstant}
\quad , 
\end {equation}
where   $r_0$ and $v_0$ are the position and the velocity
when    $t=t_0$.
The {\it second} law  is 
the conservation of energy which will be introduced 
in details in the next section.
An example is given by   the energy  conserving phase in the 
interstellar bubbles, see \cite{McCray1987}.   
The {\it third} law  of conservation 
is given by the  conservation  of momentum  flux 
which is the  rate of transfer of momentum  through a unit area
\begin{equation}
\rho(x_0)  v_0^2  A(x_0)  =
\rho(x)    v(x)^2 A(x)
\quad  ,
\label{conservazione}
\end {equation}
where  
$\rho(x)$ is the density at position $x$,
$A(x)$ is the area at position $x$ and
$v(x)$ is the velocity at  position $x$,  
see Formula A27 in \cite{deyoung}.
This  law is  useful to model the radiogalaxies 
where there is a continuous flow of matter from the central region 
to the periphery, see \cite{Zaninetti2015d}.
The {\it fourth} law  of conservation 
is given by the  conservation  of energy flux 
which is the  rate of transfer of energy through a unit area
\begin{equation}
\frac{1}{2} \rho(x_0)  v_0^3   A(x_0)  =
\frac{1}{2} \rho(x  )   v(x)^3 A(x)
\label{conservazioneenergy}
\end {equation}
where 
$\rho(x)$ is the density at position $x$,
$A(x)$ is the area at position $x$ and
$v(x)$ is the velocity at  position $x$,  
see Formula A28 in \cite{deyoung}.
This  law is  useful to model the astrophysical jets,
see \cite{Zaninetti2016e}

\section{Energy conservation}
\label{section_energy}
The conservation of kinetic energy in
spherical coordinates
within the framework of the thin
layer approximation  
when the thermal effects are negligible 
is  
\begin{equation}
\frac{1}{2} M_0(r_0) \,v_0^2 = \frac{1}{2}M(r) \,v^2 
\quad ,
\end{equation}
where $M_0(r_0)$ and $M(r)$ are the swept masses at $r_0$ and $r$,
while $v_0$ and $v$ are the velocities of the thin layer at $r_0$ and $r$.
The above conservation law, when written as a differential
equation, is  
\begin{equation}
\frac{1}{2}\,M \left( r \right)  \left( {\frac {\rm d}{{\rm d}t}}r \left( t
 \right)  \right) ^{2}-\frac{1}{2}\,{\it M_0}\,{{\it v_0}}^{2}=0
\quad  .
\end{equation}
The velocity as a function of the momentary radius is 
\begin{equation}
v(r;r_0,v_0) =
{\frac {{r_{{0}}}^{3/2}v_{{0}}}{{r}^{3/2}}}
\quad .
\end{equation}
In the following, the case of constant density
as well as 7 profiles of decreasing density will be considered.

\subsection{Medium with constant density}

\label{section_constant}
When the ISM is considered to have constant density,  
the analytical solution 
for the trajectory when the energy is conserved is 
\begin{equation} 
r(t;t_0,r_0,v_0) =
\frac{1}{2}\,{2}^{3/5}{r_{{0}}}^{3/5} \left(  \left( 5\,t-5\,t_{{0}} \right) v
_{{0}}+2\,r_{{0}} \right) ^{2/5}
\label{energy_rtconstant}
\quad , 
\end {equation}
which has  the asymptotic behaviour $ r_a(t;t_0,r_0,v_0)$,
\begin{equation} 
r_a(t;t_0,r_0,v_0) \sim
\frac{1}{2}\,{\frac {{2}^{3/5}{r_{{0}}}^{3/5}{5}^{2/5}{v_{{0}}}^{2/5}}{
 \left( {t}^{-1} \right) ^{2/5}}}
+ \frac{1}{25}\,{\frac {{2}^{3/5}{r_{{0}}}^{3/
5}{5}^{2/5} \left( -5\,t_{{0}}v_{{0}}+2\,r_{{0}} \right)  \left( {t}^{
-1} \right) ^{3/5}}{{v_{{0}}}^{3/5}}}
\quad .
\end{equation}
The velocity as function of the radius is
\begin{equation}
v(r;r_0,v_0) =
{\frac {{r_{{0}}}^{3/2}v_{{0}}}{{r}^{3/2}}}
\quad ,
\label{vfirst}
\end{equation}
and the velocity as a function of time is 
\begin{equation} 
v(t;t_0,r_0,v_0) =
{\frac {{2}^{3/5}{r_{{0}}}^{3/5}v_{{0}}}{ \left(  \left( 5\,t-5\,t_{{0
}} \right) v_{{0}}+2\,r_{{0}} \right) ^{3/5}}}
\label{energy_vtconstant}
\quad , 
\end {equation}
where   $r_0$ and $v_0$ are the position and the velocity
when    $t=t_0$.

\subsection{Constant density and back reaction}
The radiative losses per unit length  
are assumed to be  proportional to the   flux of momentum 
\begin{equation}
- \epsilon \rho_s v^2 4\,\pi r^2 
\quad ,
\end{equation}
where $\epsilon$ is a constant and $rho_s$ is density
in the thin advancing layer which is $4\,\rho$.
Inserting in the above equation  the  velocity to first order
as  given by equation~(\ref{vfirst}) 
the radiative losses, $Q(r;r_0,v_0,\epsilon)$, are
\begin{equation}
Q(r;r_0,v_0,\epsilon)=
-16\,{\frac {\epsilon\,\rho\,{r_{{0}}}^{3}{v_{{0}}}^{2}\pi}{r}}
\quad .
\label{lossesclassical}
\end{equation}
The sum of the radiative  losses between $r_0$ and $r$ 
is given by the following integral, $L$,
\begin{equation}
L(r;r_0,v_0,\epsilon)=\int_{r_0}^r  Q(r;r_0,v_0,\epsilon) dr
=
-16\,\epsilon\,\rho\,{r_{{0}}}^{3}{v_{{0}}}^{2}\pi\,\ln  \left( r
 \right) +16\,\epsilon\,\rho\,{r_{{0}}}^{3}{v_{{0}}}^{2}\pi\,\ln 
 \left( r_{{0}} \right) 
\quad .
\label{classiclosses}
\end{equation}
The  conservation of energy  in  presence  
of  the back reaction due to the radiative losses
is 
\begin{eqnarray}
2/3\,\rho\,\pi\,{r}^{3}{v}^{2}+16\,\epsilon\,\rho\,{r_{{0}}}^{3}{v_{{0
}}}^{2}\pi\,\ln  \left( r \right) -16\,\epsilon\,\rho\,{r_{{0}}}^{3}{v
_{{0}}}^{2}\pi\,\ln  \left( r_{{0}} \right) =2/3\,\rho\,\pi\,{r_{{0}}}
^{3}{v_{{0}}}^{2}
\quad . 
\label{eqnenergyback}
\end{eqnarray}
The  analytical solution for the velocity to 
second order, $v_c(r;r_0,c_0,\epsilon)$, 
is
\begin{equation}
v_c(r;r_0,v_0,\epsilon)=
{\frac {{r_{{0}}}^{3/2}\sqrt {-24\,\ln  \left( r \right) \epsilon+24\,
\ln  \left( r_{{0}} \right) \epsilon+1}v_{{0}}}{{r}^{3/2}}}
\label{vcorrected}
\quad .
\end{equation}
The  inclusion  of back reaction  allows the evaluation of the 
SRS's maximum length $r_{back}(r_0,\epsilon)$ ,  which can be derived 
imposing to zero the above velocity.
\begin{equation}
r_{back}(r_0,\epsilon) 
= 
{{\rm e}^{1/24\,{\frac {24\,\ln  \left( r_{{0}} \right) \epsilon+1}{
\epsilon}}}}
\quad .
\end{equation}

\subsection{Medium with an hyperbolic profile of  density }

\label{section_hyperbolic}
We  assume that the medium 
around the SN
scales with the piecewise dependence
\begin{equation}
 \rho (r;r_0)  = \{ \begin{array}{ll}
            \rho_c                      & \mbox {if $r \leq r_0 $ } \\
            \rho_c (\frac{r_0}{r})    & \mbox {if $r >     r_0 $.}
            \end{array}
\label{piecewiseinverse},
\end{equation}
where 
$\rho_c$ is the density at $r=0$
and  
$r_0$ is the radius after which the density 
starts to decrease.
The mass swept, $M_0$,
in the interval [0,$r_0$]
is
\begin{equation}
M_0(\rho_c,r_0) =
\frac{4}{3}\,\rho_{{c}}\pi \,{r_{{0}}}^{3}
\quad .
\nonumber
\end{equation}
The total mass swept, $ M(r;r_0,\rho_c) $,
in the interval [0,r]
is
\begin{equation}
M (r;r_0,\rho_c)=
-\frac{2}{3}\,\rho_{{c}}\pi\,{r_{{0}}}^{3}+2\,\rho_{{c}}r_{{0}}{r}^{2}\pi
\quad .
\nonumber
\label{masshyperbolic}
\end{equation}
The  application of energy conservation gives 
the velocity as a function of the radius:
\begin{equation}
v(r;r_0,v_0)=
2\,{\frac {v_{{0}}r_{{0}}}{\sqrt {6\,{r}^{2}-2\,{r_{{0}}}^{2}}}}
\quad  .
\end{equation}
Separation of variables 
followed by integration
gives
\begin{eqnarray}
\frac{1}{12}\,{\frac {r_{{0}}\sqrt {6}\ln  \left( \sqrt {2}+\sqrt {3} \right) 
}{v_{{0}}}}-\frac{1}{12}\,{\frac {r_{{0}}\sqrt {6}\ln  \left( r\sqrt {2}\sqrt 
{3}+\sqrt {6\,{r}^{2}-2\,{r_{{0}}}^{2}} \right) }{v_{{0}}}}+\frac{1}{24}\,{
\frac {r_{{0}}\sqrt {6}\ln  \left( 2 \right) }{v_{{0}}}}
\nonumber \\ 
+\frac{1}{12}\,{\frac 
{r_{{0}}\sqrt {6}\ln  \left( r_{{0}} \right) }{v_{{0}}}}+\frac{1}{4}\,{\frac {
r\sqrt {6\,{r}^{2}-2\,{r_{{0}}}^{2}}}{v_{{0}}r_{{0}}}}-\frac{1}{2}\,{\frac {r_
{{0}}}{v_{{0}}}}
=t-t0
\quad . 
\label{trrelation}
\end{eqnarray}
In this equation it is not possible  to extract the 
radius as a function of time, and therefore a numerical procedure 
is adopted in order to derive the trajectory.

\subsection{Medium with an inverse square profile for the density }

\label{section_invsquare}
We now  assume that the medium 
around the SN
scales with the piecewise dependence
(which avoids a pole at $r=0$)
\begin{equation}
 \rho (r;r_0)  = \{ \begin{array}{ll}
            \rho_c                      & \mbox {if $r \leq r_0 $ } \\
            \rho_c (\frac{r_0}{r})^2    & \mbox {if $r >     r_0 $.}
            \end{array}
\label{piecewiseinvsquare},
\end{equation}
where 
$\rho_c$ is the density at $r=0$
and  
$r_0$ is the radius after which the density 
starts to decrease.

The total mass swept, $ M(r;r_0,\rho_c) $,
in the interval [0,r]
is
\begin{eqnarray}
M (r;r_0,\rho_c)=
-\frac{8}{3}\,\rho_{{c}}\pi\,{r_{{0}}}^{3}+4\,\rho_{{c}}{r_{{0}}}^{2}\pi\,r
+ \frac{4}{3}\,\rho_{{c}}\pi \,{r_{{0}}}^{3}
\quad .
\nonumber
\label{massinversesquare}
\end{eqnarray}
Applying the conservation of energy,
the velocity as a function of the radius is  
\begin{equation}
v(r;r_0,v_0)=
-{\frac {\sqrt {- \left( 2\,r_{{0}}-3\,r \right) r_{{0}}}v_{{0}}}{2\,r
_{{0}}-3\,r}}
\quad  .
\end{equation}
The trajectory, i.e. the radius as a function of time,  is
\begin{equation}
r(t;t_0,r_0,v_0)= 
\frac{1}{6}\,\sqrt [3]{2}\sqrt [3]{r_{{0}}} \left(  \left( 9\,t-9\,t_{{0}}
 \right) v_{{0}}+2\,r_{{0}} \right) ^{2/3}+\frac{2}{3}\,r_{{0}}
\quad,
\label{rtinversesquare}
\end{equation}
which has the asymptotic behavior, $r_a(t;t_0,r_0,v_0)$,
\begin{equation}
r_a(t;t_0,r_0,v_0)
\sim
\frac{1}{6}\,{\frac {\sqrt [3]{2}\sqrt [3]{r_{{0}}}{9}^{2/3}{v_{{0}}}^{2/3}}{
 \left( {t}^{-1} \right) ^{2/3}}}+\frac{2}{3}\,r_{{0}}+{\frac {\sqrt [3]{2}
\sqrt [3]{r_{{0}}}{9}^{2/3} \left( -9\,t_{{0}}v_{{0}}+2\,r_{{0}}
 \right) \sqrt [3]{{t}^{-1}}}{81\,\sqrt [3]{v_{{0}}}}}
\quad.
\end{equation}
The velocity as a function of 
 time is 
\begin{equation} 
v(t;t_0,r_0,v_0) =
{\frac {\sqrt [3]{2}\sqrt [3]{r_{{0}}}v_{{0}}}{\sqrt [3]{ \left( 9\,t-
9\,t_{{0}} \right) v_{{0}}+2\,r_{{0}}}}}
\quad .
\end{equation}

\subsection{Medium with a  power law profile for the density }

\label{section_powerlaw}
We now  assume that the medium 
around the SN
scales 
as
\begin{equation}
 \rho (r;r_0)  = \{ \begin{array}{ll}
            \rho_c                      & \mbox {if $r \leq r_0 $ } \\
            \rho_c (\frac{r_0}{r})^{\alpha}    & \mbox {if $r >     r_0 $.}
            \end{array}
\label{piecewisealpha},
\end{equation}
where 
$\rho_c$ is the density at $r=0$,
$r_0$ is the radius after which the density 
starts to decrease
and
$\alpha >0$.

The total mass swept, $ M(r;r_0,\rho_c,\alpha) $,
in the interval [0,r]
is
\begin{eqnarray}
M (r;r_0,\rho_c,\alpha)=
\frac{4}{3}\,\rho_{{c}}\pi\,{r_{{0}}}^{3}-4\,{\frac {{r}^{3}\rho_{{c}}\pi}{
\alpha-3} \left( {\frac {r_{{0}}}{r}} \right) ^{\alpha}}+4\,{\frac {
\rho_{{c}}\pi\,{r_{{0}}}^{3}}{\alpha-3}}
\quad .
\nonumber
\label{massinversepowerlaw}
\end{eqnarray}
The  application of energy conservation 
gives the differential  equation
\begin{equation}
\frac 
{1}
{
3\,\alpha-9
}
\Bigg (
-2\,\rho_{{c}}\pi\, \left( 3\,{r}^{3} \left( {\frac {r_{{0}}}{r}}
 \right) ^{\alpha}-{r_{{0}}}^{3}\alpha \right)  \left( {\frac {\rm d}{
{\rm d}t}}r \left( t \right)  \right) ^{2}
\Bigg )
=
\frac{2}{3}\,\rho_{{c}}\pi\,{r_{{0}}}^{3}{v_{{0}}}^{2}
\quad.
\label{eqndiffalpha}
\end{equation}
The velocity as a function of the radius
is 
\begin{equation}
v(r;r_0,v_0,\alpha)=
{\frac {\sqrt {- \left( -{r_{{0}}}^{3}\alpha+3\,{r}^{3-\alpha}{r_{{0}}
}^{\alpha} \right) r_{{0}} \left( \alpha-3 \right) }v_{{0}}r_{{0}}}{-{
r_{{0}}}^{3}\alpha+3\,{r}^{3-\alpha}{r_{{0}}}^{\alpha}}}
\quad .
\label{velocityradiusalpha}
\end{equation}
There is no analytical solution for the trajectory, and 
therefore we have implemented a numerical procedure. 
The first approximation for the trajectory 
is obtained by a series solution of Equation (\ref{eqndiffalpha})
to fourth order,
\begin{equation}
r(t;r_0,v_0,t_0,\alpha) \approx
r_{{0}}+v_{{0}} \left( t-t_{{0}} \right) -\frac{3}{4}\,{\frac {{v_{{0}}}^{2}
 \left( t-t_{{0}} \right) ^{2}}{r_{{0}}}}
+\frac{1}{4}\,{\frac {{v_{{0}}}^{3}
 \left( \alpha+4 \right)  \left( t-t_{{0}} \right) ^{3}}{{r_{{0}}}^{2}
}}
\quad  .
\label{rttayloralpha}
\end{equation}
The second approximation for the trajectory
is found by first deriving   an asymptotic expansion  of 
Equation (\ref{velocityradiusalpha}), namely 
\begin{equation}
v(r;r_0,v_0,\alpha) \sim
\frac{1}{3}\,{\frac {v_{{0}}r_{{0}}\sqrt {3}\sqrt {{r_{{0}}}^{\alpha+1}
 \left( 3-\alpha \right) }}{{r_{{0}}}^{\alpha}\sqrt { \left( {r}^{-1}
 \right) ^{\alpha-3}}}}
\quad .
\end{equation}
Then, the asymptotic approximate trajectory
turns out to be 
\begin{eqnarray}
r(t;r_0,v_0,t_0,\alpha) \sim
{12}^{ \left( \alpha-5 \right) ^{-1}}{r_{{0}}}^{{\frac {\alpha-3}{
\alpha-5}}} \times  
\nonumber \\
\left( -4\,r_{{0}}v_{{0}} \left( \alpha-5 \right)  \left( 
t-t_{{0}} \right) \sqrt {9-3\,\alpha}- \left( \alpha-3 \right) 
 \left( \alpha-5 \right) ^{2} \left( t-t_{{0}} \right) ^{2}{v_{{0}}}^{
2}+12\,{r_{{0}}}^{2} \right) ^{- \left( \alpha-5 \right) ^{-1}}
\quad .
\label{rtalphaasympt}
\end{eqnarray}

\subsection{Medium with an exponential profile for the density }

\label{section_exponential}
We  assume that the medium 
around the SN
scales with the piecewise dependence
\begin{equation}
 \rho (r;r_0)  = \{ \begin{array}{ll}
            \rho_c                        & \mbox {if $r \leq r_0 $ } \\
            \rho_c (\exp{-\frac{r}{b}})    & \mbox {if $r >     r_0 $.}
            \end{array}
\label{piecewiseexp},
\end{equation}
where 
$\rho_c$ is the density at $r=0$
and  
$r_0$ is the radius after which the density 
starts to decrease.
The total mass swept, $ M(r;r_0,\rho_c) $,
in the interval [0,r]
is
\begin{equation}
M (r;r_0,\rho_c,b)=
\frac{4}{3}\,\rho_{{c}}\pi\,{r_{{0}}}^{3}-4\,b \left( 2\,{b}^{2}+2\,br+{r}^{2}
 \right) \rho_{{c}}{{\rm e}^{-{\frac {r}{b}}}}\pi+4\,b \left( 2\,{b}^{
2}+2\,br_{{0}}+{r_{{0}}}^{2} \right) \rho_{{c}}{{\rm e}^{-{\frac {r_{{0
}}}{b}}}}\pi
\quad .
\nonumber
\label{massexponential}
\end{equation}
The  application of energy conservation 
gives the differential  equation
\begin{eqnarray}
-2\, \left( {\frac {\rm d}{{\rm d}t}}r \left( t \right)  \right) ^{2}
\rho_{{c}} 
\Big( 6\,{b}^{3}{{\rm e}^{-{\frac {r}{b}}}}+6\,{b}^{2}r{
{\rm e}^{-{\frac {r}{b}}}}+3\,b{r}^{2}{{\rm e}^{-{\frac {r}{b}}}}-6\,{
b}^{3}{{\rm e}^{-{\frac {r_{{0}}}{b}}}}
\nonumber \\
-6\,{b}^{2}{{\rm e}^{-{\frac {r
_{{0}}}{b}}}}r_{{0}}-3\,b{{\rm e}^{-{\frac {r_{{0}}}{b}}}}{r_{{0}}}^{2
}-{r_{{0}}}^{3} \Big) \pi
=
\frac{2}{3}\,\rho_{{c}}\pi\,{r_{{0}}}^{3}{v_{{0}}}^{2}
\quad.
\label{eqndiffexp}
\end{eqnarray}
The velocity as a function of the radius
is 
\begin{equation}
v(r;r_0,v_0,b)=
\frac{N}{D}
\quad ,
\label{velocityradiusexp}
\end{equation}
where 
\begin{equation}
N=-\sqrt {-6\,r_{{0}} \left(  \left( -{b}^{3}-{b}^{2}r_{{0}}-\frac{1}{2}\,b{r_{{0
}}}^{2} \right) {{\rm e}^{-{\frac {r_{{0}}}{b}}}}+b \left( {b}^{2}+br+
\frac{1}{2}\,{r}^{2} \right) {{\rm e}^{-{\frac {r}{b}}}}-1/6\,{r_{{0}}}^{3}
 \right) }v_{{0}}r_{{0}}
\quad ,
\end{equation}
and  
\begin{equation}
D=
\left( -6\,{b}^{3}-6\,{b}^{2}r_{{0}}-3\,b{r_{{0}}}^{2} \right) {
{\rm e}^{-{\frac {r_{{0}}}{b}}}}+ \left( 6\,{b}^{3}+6\,{b}^{2}r+3\,b{r
}^{2} \right) {{\rm e}^{-{\frac {r}{b}}}}-{r_{{0}}}^{3}
\quad .
\end{equation}
There is no analytical solution for the trajectory,
and therefore we 
present
a series solution of Equation (\ref{eqndiffexp})
to fourth order:
\begin{eqnarray}
r(t;r_0,v_0,t_0,b) \approx
r_{{0}}+ \left( t-t_{{0}} \right) v_{{0}}-\frac{3}{4}\,{\frac {{v_{{0}}}^{2}
 \left( t-t_{{0}} \right) ^{2}}{r_{{0}}}{{\rm e}^{-{\frac {r_{{0}}}{b}
}}}}
\nonumber \\
+ \frac{1}{4}\,{\frac {{v_{{0}}}^{3} \left( t-t_{{0}} \right) ^{3}}{b{r_{{0
}}}^{2}}{{\rm e}^{-{\frac {r_{{0}}}{b}}}} \left( 6\,b{{\rm e}^{-{
\frac {r_{{0}}}{b}}}}-2\,b+r_{{0}} \right) }
\quad  .
\label{rttaylorexp}
\end{eqnarray}

\subsection{Medium with a Gaussian profile for the density }

\label{section_gaussian}
We  assume that the medium 
around the SN
scales with the piecewise dependence
\begin{equation}
 \rho (r;r_0,b)  = \{ \begin{array}{ll}
            \rho_c                        & \mbox {if $r \leq r_0 $ } \\
            \rho_c (\exp{-(\frac{r}{b})^2})    & \mbox {if $r >     r_0 $.}
            \end{array}
\label{piecewisegaussian},
\end{equation}
where 
$\rho_c$ is the density at $r=0$
and  
$r_0$ is the radius after which the density 
starts to decrease.
The total mass swept, $ M(r;r_0,\rho_c) $,
in the interval [0,r]
is
\begin{eqnarray}
M (r;r_0,\rho_c,b)=
\frac{4}{3}\,\rho_{{c}}\pi\,{r_{{0}}}^{3}
\nonumber \\
+4\,\rho_{{c}}\pi\, \left( -\frac{1}{2}\,{
{\rm e}^{-{\frac {{r}^{2}}{{b}^{2}}}}}r{b}^{2}
+\frac{1}{4}\,{b}^{3}\sqrt {\pi}
{\rm erf} \left({\frac {r}{b}}\right) \right) -4\,\rho_{{c}}\pi\,
 \left( -\frac{1}{2}\,{{\rm e}^{-{\frac {{r_{{0}}}^{2}}{{b}^{2}}}}}r_{{0}}{b}^
{2}+\frac{1}{4}\,{b}^{3}\sqrt {\pi}{\rm erf} \left({\frac {r_{{0}}}{b}}\right)
 \right) 
\label{massegaussian}
\quad , 
\end{eqnarray}
where ${\rm erf(x)}$ is the error function, defined by
\begin{equation}
\mathop{\mathrm{erf}\/}\nolimits
(x)=\frac{2}{\sqrt{\pi}}\int_{0}^{x}e^{-t^{2}}dt
\quad ,
\end{equation}
see \cite{NIST2010}.

The  differential  equation  when the  energy is conserved 
is  
\begin{eqnarray}
-\frac{1}{6}  \Big ( {\frac {\rm d}{{\rm d}t}}r   ( t   )  \Big  ) ^{2}\pi
\,\rho_{{c}} \Big  ( -3\,{b}^{3}\sqrt {\pi}{\rm erf}   ({\frac {r
   ( t   ) }{b}}  )+3\,{b}^{3}\sqrt {\pi}{\rm erf}   ({
\frac {r_{{0}}}{b}}  )
\nonumber  \\
+6\,{{\rm e}^{-{\frac {   ( r   ( t
   )    ) ^{2}}{{b}^{2}}}}}r   ( t   ) {b}^{2}-6\,{
{\rm e}^{-{\frac {{r_{{0}}}^{2}}{{b}^{2}}}}}r_{{0}}{b}^{2}-4\,{r_{{0}}
}^{3}  \Big ) 
=
\frac{2}{3}\,\rho_{{c}}\pi\,{r_{{0}}}^{3}{v_{{0}}}^{2}
\quad  .
\label{eqndiffgaussian}
\end{eqnarray}
In the absence of an analytical solution for this differential equation,
we present an approximation using the fourth order Taylor series:
\begin{eqnarray}
r(t;r_0,v_0,t_0,b) \approx
r_{{0}}+v_{{0}} \left( t-t_{{0}} \right) 
\nonumber  \\
-\frac{3}{4}\,{\frac {{v_{{0}}}^{2}
 \left( t-t_{{0}} \right) ^{2}}{r_{{0}}}{{\rm e}^{-{\frac {{r_{{0}}}^{
2}}{{b}^{2}}}}}}+\frac{1}{2}\,{\frac {{v_{{0}}}^{3} \left( t-t_{{0}} \right) ^
{3}}{{r_{{0}}}^{2}{b}^{2}}{{\rm e}^{-{\frac {{r_{{0}}}^{2}}{{b}^{2}}}}
} \left( 3\,{b}^{2}{{\rm e}^{-{\frac {{r_{{0}}}^{2}}{{b}^{2}}}}}-{b}^{
2}+{r_{{0}}}^{2} \right) }
\quad  .
\label{rttaylorgauss}
\end{eqnarray}

\subsection{Autogravitating medium}

\label{section_sech2}
We  assume that the medium 
around the SN
scales with the piecewise dependence
\begin{equation}
 \rho (r;r_0,b)  = \{ \begin{array}{ll}
            \rho_c                          & \mbox {if $r \leq r_0 $ } \\
            \rho_c (sech^2 (\frac{r}{2\,b}))& \mbox {if $r >     r_0 $.}
            \end{array}
\label{piecewisech2},
\end{equation}
where 
$\rho_c$ is the density at $r=0$,
$r_0$ is the radius after which the density 
starts to decrease
and  $sech$ is the hyperbolic secant  
(\cite{Spitzer1942,Rohlfs1977,Bertin2000,Padmanabhan_III_2002}).

The total mass swept, $ M(r;r_0,b,\rho_c) $,
in the interval [0,r]
is
\begin{eqnarray}
M (r;r_0,\rho_c,b)=
\frac{4}{3}\,\rho_{{c}}\pi\,{r_{{0}}}^{3}-16\,{\rho_{{c}}\pi\,{r}^{2}b \left( 
1+{{\rm e}^{{\frac {r}{b}}}} \right) ^{-1}}-32\,\rho_{{c}}\pi\,{b}^{2}
r\ln  \left( 1+{{\rm e}^{{\frac {r}{b}}}} \right) 
\nonumber \\
-32\,\rho_{{c}}\pi\,
{b}^{3}{\it polylog} \left( 2,-{{\rm e}^{{\frac {r}{b}}}} \right) +16
\,\rho_{{c}}\pi\,{r}^{2}b+16\,{\rho_{{c}}\pi\,{r_{{0}}}^{2}b \left( 1+
{{\rm e}^{{\frac {r_{{0}}}{b}}}} \right) ^{-1}}
\nonumber \\
+32\,\rho_{{c}}\pi\,{b}
^{2}r_{{0}}\ln  \left( 1+{{\rm e}^{{\frac {r_{{0}}}{b}}}} \right) +32
\,\rho_{{c}}\pi\,{b}^{3}{\it polylog} \left( 2,-{{\rm e}^{{\frac {r_{{0
}}}{b}}}} \right) -16\,\rho_{{c}}\pi\,{r_{{0}}}^{2}b
\label{masssech2}
\quad , 
\end{eqnarray}
where the polylog operator  is
defined by
\begin{equation}
polylog(s,z) =
\mathrm{Li}_{s}\left(z\right) =\sum_{n=1}^{\infty}\frac{z^{n}}{n^{s}}
\quad 
\end{equation}
and  $\mathrm{Li}_{s}\left(z\right)$ is
a Dirichlet series.
The  differential  equation  when the  energy is conserved 
is  
\begin{equation}
\frac{ODEN}{3\, \left( 1+{{\rm e}^{{\frac {r \left( t \right) }{b}}}}
\right) 
 \left( 1+{{\rm e}^{{\frac {r_{{0}}}{b}}}} \right)
}=
\frac{2}{3}\,\rho_{{c}}\pi\,{r_{{0}}}^{3}{v_{{0}}}^{2}
\end{equation}
where 
\begin{eqnarray}
ODEN=
48\,   ( {\frac {\rm d}{{\rm d}t}}r   ( t   )    ) ^{2}
 \Big  ( -{b}^{3}   ( {{\rm e}^{{\frac {r   ( t   ) +r_{{0}}
}{b}}}}+{{\rm e}^{{\frac {r_{{0}}}{b}}}}+{{\rm e}^{{\frac {r   ( t
   ) }{b}}}}+1   ) {\it polylog}   ( 2,-{{\rm e}^{{\frac {r
   ( t   ) }{b}}}}   ) +   ( -{b}^{2}r   ( t   ) 
\ln    ( 1+{{\rm e}^{{\frac {r   ( t   ) }{b}}}}   ) 
\nonumber \\
+{b
}^{3}{\it polylog}   ( 2,-{{\rm e}^{{\frac {r_{{0}}}{b}}}}   ) +{b}^
{2}r_{{0}}\ln    ( 1+{{\rm e}^{{\frac {r_{{0}}}{b}}}}   ) +\frac{1}{2}
\,   ( r   ( t   )    ) ^{2}b-\frac{1}{2}\,{r_{{0}}}^{2}   ( 
b-\frac{1}{12}\,r_{{0}}   )    ) {{\rm e}^{{\frac {r   ( t   ) 
+r_{{0}}}{b}}}}
\nonumber \\
-{b}^{2}r   ( t   )    ( {{\rm e}^{{\frac {r_
{{0}}}{b}}}}+{{\rm e}^{{\frac {r   ( t   ) }{b}}}}+1   ) 
\ln    ( 1+{{\rm e}^{{\frac {r   ( t   ) }{b}}}}   ) +{b
}^{3}   ( {{\rm e}^{{\frac {r_{{0}}}{b}}}}+{{\rm e}^{{\frac {r
   ( t   ) }{b}}}}+1   ) {\it polylog}   ( 2,-{{\rm e}^{{
\frac {r_{{0}}}{b}}}}   ) 
\nonumber  \\
+{b}^{2}r_{{0}}   ( {{\rm e}^{{\frac 
{r_{{0}}}{b}}}}+{{\rm e}^{{\frac {r   ( t   ) }{b}}}}+1
   ) \ln    ( 1+{{\rm e}^{{\frac {r_{{0}}}{b}}}}   ) +
   ( \frac{1}{2}\,   ( r   ( t   )    ) ^{2}b+1/24\,{r_{{0}}}
^{3}   ) {{\rm e}^{{\frac {r   ( t   ) }{b}}}}-1/2\,{r_{{0}
}}^{2}   (    ( b-\frac{1}{12}\,r_{{0}}   ) {{\rm e}^{{\frac {r_{{0}
}}{b}}}}
\nonumber  \\
-\frac{1}{12}\,r_{{0}}   ) \Big   ) \rho_{{c}}\pi
\quad .
\end{eqnarray}
The velocity as a function of the radius
is
\begin{equation}
v (r;r_0,b)
=
\frac
{
{r_{{0}}}^{\frac{3}{2}}\sqrt {{{\rm e}^{{\frac {r_{{0}}+r}{b}}}}+{{\rm e}^{{
\frac {r_{{0}}}{b}}}}+{{\rm e}^{{\frac {r}{b}}}}+1}v_{{0}}
}
{
VELD
}
\,
\end{equation}
where 
\begin{eqnarray}
VELD=
\Bigg (24\,{b}^{3}{\it polylog} \left( 2,-{{\rm e}^{{\frac {r_{{0}}}{b
}}}} \right) {{\rm e}^{{\frac {r_{{0}}+r}{b}}}}+24\,{b}^{3}{{\rm e}^{{
\frac {r}{b}}}}{\it polylog} \left( 2,-{{\rm e}^{{\frac {r_{{0}}}{b}}}
} \right) +24\,{b}^{3}{{\rm e}^{{\frac {r_{{0}}}{b}}}}{\it polylog}
 \left( 2,-{{\rm e}^{{\frac {r_{{0}}}{b}}}} \right) 
\nonumber \\
-24\,{b}^{3}{\it 
polylog} \left( 2,-{{\rm e}^{{\frac {r}{b}}}} \right) {{\rm e}^{{
\frac {r_{{0}}+r}{b}}}}-24\,{b}^{3}{{\rm e}^{{\frac {r}{b}}}}{\it 
polylog} \left( 2,-{{\rm e}^{{\frac {r}{b}}}} \right) -24\,{b}^{3}{
{\rm e}^{{\frac {r_{{0}}}{b}}}}{\it polylog} \left( 2,-{{\rm e}^{{
\frac {r}{b}}}} \right) 
\nonumber \\
+24\,\ln  \left( 1+{{\rm e}^{{\frac {r_{{0}}}{
b}}}} \right) {{\rm e}^{{\frac {r_{{0}}+r}{b}}}}{b}^{2}r_{{0}}+24\,{b}
^{2}r_{{0}}{{\rm e}^{{\frac {r}{b}}}}\ln  \left( 1+{{\rm e}^{{\frac {r
_{{0}}}{b}}}} \right) +24\,{b}^{2}r_{{0}}{{\rm e}^{{\frac {r_{{0}}}{b}
}}}\ln  \left( 1+{{\rm e}^{{\frac {r_{{0}}}{b}}}} \right) 
\nonumber \\
-24\,\ln 
 \left( 1+{{\rm e}^{{\frac {r}{b}}}} \right) {{\rm e}^{{\frac {r_{{0}}
+r}{b}}}}{b}^{2}r-24\,{b}^{2}r{{\rm e}^{{\frac {r}{b}}}}\ln  \left( 1+
{{\rm e}^{{\frac {r}{b}}}} \right) -24\,{b}^{2}r{{\rm e}^{{\frac {r_{{0
}}}{b}}}}\ln  \left( 1+{{\rm e}^{{\frac {r}{b}}}} \right) 
\nonumber \\
+24\,{b}^{3}
{\it polylog} \left( 2,-{{\rm e}^{{\frac {r_{{0}}}{b}}}} \right) -24\,
{b}^{3}{\it polylog} \left( 2,-{{\rm e}^{{\frac {r}{b}}}} \right) +24
\,{b}^{2}r_{{0}}\ln  \left( 1+{{\rm e}^{{\frac {r_{{0}}}{b}}}}
 \right) -24\,{b}^{2}r\ln  \left( 1+{{\rm e}^{{\frac {r}{b}}}}
 \right) 
\nonumber  \\
+12\,{{\rm e}^{{\frac {r_{{0}}+r}{b}}}}b{r}^{2}-12\,{{\rm e}^
{{\frac {r_{{0}}+r}{b}}}}b{r_{{0}}}^{2}+{{\rm e}^{{\frac {r_{{0}}+r}{b
}}}}{r_{{0}}}^{3}+12\,b{r}^{2}{{\rm e}^{{\frac {r}{b}}}}+{r_{{0}}}^{3}
{{\rm e}^{{\frac {r}{b}}}}-12\,b{r_{{0}}}^{2}{{\rm e}^{{\frac {r_{{0}}
}{b}}}}+{r_{{0}}}^{3}{{\rm e}^{{\frac {r_{{0}}}{b}}}}+{r_{{0}}}^{3}
\Bigg )^{1/2}
\quad .
\end{eqnarray}
In the absence of an analytical solution for this differential equation,
we present the approximation arising from the fourth order Taylor series:
\begin{eqnarray}
r(t;r_0,v_0,t_0,b) \approx
r_{{0}}+v_{{0}} \left( t-t_{{0}} \right) 
\nonumber  \\
+3\,{\frac {{v_{{0}}}^{2}
 \left( t-t_{{0}} \right) ^{2}}{{r_{{0}}}^{2}} \left( 2\, \left( {
{\rm e}^{{\frac {r_{{0}}}{b}}}} \right) ^{2}b- \left( {{\rm e}^{{
\frac {r_{{0}}}{b}}}} \right) ^{2}r_{{0}}-{{\rm e}^{{\frac {r_{{0}}}{b
}}}}r_{{0}}-2\,b{{\rm e}^{2\,{\frac {r_{{0}}}{b}}}} \right)  \left( 1+
{{\rm e}^{{\frac {r_{{0}}}{b}}}} \right) ^{-1} \left( {{\rm e}^{2\,{
\frac {r_{{0}}}{b}}}}+2\,{{\rm e}^{{\frac {r_{{0}}}{b}}}}+1 \right) ^{
-1}}
\nonumber  \\
+{\frac {{v_{{0}}}^{3} \left( t-t_{{0}} \right) ^{3}}{{r_{{0}}}^{2
}b} \left( -2\,b{{\rm e}^{2\,{\frac {r_{{0}}}{b}}}}+{{\rm e}^{2\,{
\frac {r_{{0}}}{b}}}}r_{{0}}+20\,b{{\rm e}^{{\frac {r_{{0}}}{b}}}}-2\,
b-r_{{0}} \right) {{\rm e}^{{\frac {r_{{0}}}{b}}}} \left( 1+{{\rm e}^{
{\frac {r_{{0}}}{b}}}} \right) ^{-4}}
\quad  .
\label{rttaylorsech2}
\end{eqnarray}

\subsection{Medium with an NFW profile}

\label{section_nfw}

We  assume that the medium 
around the SN
scales with the
Navarro--Frenk--White (NFW) distribution  
as follows:
\begin{equation}
 \rho (r;r_0,b)  = \Bigg \{ \begin{array}{ll}
            \rho_c                          & \mbox {if $r \leq r_0 $ } \\
            \frac
             {
              \rho_{{c}}r_{{0}} \left( b+r_{{0}} \right) ^{2}
             }
             {
              r \left( b+r \right) ^{2}
             }
                 & \mbox {if $r >     r_0 $}
            \end{array}
\label{piecewisenfw},
\end{equation}
where 
$\rho_c$ is the density at $r=0$,
and $r_0$ is the radius after which the density 
starts to decrease,  
see \cite{Navarro1996}.  
The total mass swept, $ M(r;r_0,b,\rho_c) $,
in the interval [0,r]
is
\begin{eqnarray}
M (r;r_0,\rho_c,b)=
\frac{4}{3}\,\rho_{{c}}\pi\,{r_{{0}}}^{3}+4\,\rho_{{c}}r_{{0}}\pi\,\ln 
 \left( b+r \right) {b}^{2}+8\,\rho_{{c}}{r_{{0}}}^{2}\pi\,\ln 
 \left( b+r \right) b+4\,\rho_{{c}}{r_{{0}}}^{3}\pi\,\ln  \left( b+r
 \right)
\nonumber \\
 +4\,{\frac {\rho_{{c}}r_{{0}}\pi\,{b}^{3}}{b+r}}+8\,{\frac {
\rho_{{c}}{r_{{0}}}^{2}\pi\,{b}^{2}}{b+r}}+4\,{\frac {\rho_{{c}}\pi\,{
r_{{0}}}^{3}b}{b+r}}-4\,\rho_{{c}}r_{{0}}\pi\,\ln  \left( b+r_{{0}}
 \right) {b}^{2}-8\,\rho_{{c}}{r_{{0}}}^{2}\pi\,\ln  \left( b+r_{{0}}
 \right) b
\nonumber \\
-4\,\rho_{{c}}{r_{{0}}}^{3}\pi\,\ln  \left( b+r_{{0}}
 \right) -4\,{\frac {\rho_{{c}}r_{{0}}\pi\,{b}^{3}}{b+r_{{0}}}}-8\,{
\frac {\rho_{{c}}{r_{{0}}}^{2}\pi\,{b}^{2}}{b+r_{{0}}}}-4\,{\frac {
\rho_{{c}}\pi\,{r_{{0}}}^{3}b}{b+r_{{0}}}}
\quad .
\end{eqnarray}
The  differential  equation  when the  energy is conserved 
for an NFW profile 
is  
\begin{equation}
\frac{ODENN}
{
3\,b+3\,r \left( t \right)
}
=
\frac{2}{3}\,\rho_{{c}}\pi\,{r_{{0}}}^{3}{v_{{0}}}^{2}
\end{equation}
where 
\begin{eqnarray}
ODENN=
-2\,r_{{0}}\rho_{{c}} \Big ( 3\,\ln  \left( b+r_{{0}} \right) r
 \left( t \right) {b}^{2}+6\,\ln  \left( b+r_{{0}} \right) r \left( t
 \right) br_{{0}}+3\,\ln  \left( b+r_{{0}} \right) r \left( t \right) 
{r_{{0}}}^{2}
\nonumber  \\
+3\,{b}^{3}\ln  \left( b+r_{{0}} \right) +6\,{b}^{2}r_{{0
}}\ln  \left( b+r_{{0}} \right) +3\,b{r_{{0}}}^{2}\ln  \left( b+r_{{0}
} \right) -3\,\ln  \left( b+r \left( t \right)  \right) r \left( t
 \right) {b}^{2}
\nonumber \\
-6\,\ln  \left( b+r \left( t \right)  \right) r
 \left( t \right) br_{{0}}-3\,\ln  \left( b+r \left( t \right) 
 \right) r \left( t \right) {r_{{0}}}^{2}-3\,{b}^{3}\ln  \left( b+r
 \left( t \right)  \right) -6\,\ln  \left( b+r \left( t \right) 
 \right) {b}^{2}r_{{0}}
\nonumber  \\
-3\,\ln  \left( b+r \left( t \right)  \right) b
{r_{{0}}}^{2}+3\,r \left( t \right) {b}^{2}+3\,r \left( t \right) br_{
{0}}-r \left( t \right) {r_{{0}}}^{2}-3\,r_{{0}}{b}^{2}-4\,{r_{{0}}}^{
2}b \Big ) \pi\, \left( {\frac {\rm d}{{\rm d}t}}r \left( t \right) 
 \right) ^{2}
\quad .
\end{eqnarray}
The velocity as a function of the radius
is
\begin{equation}
v (r;r_0,b)
=
\frac
{
\sqrt {b+r}v_{{0}}r_{{0}}
}
{
VELDD
}
\,
\end{equation}
where 
\begin{eqnarray}
VELDD=
\Big 
(3\,{b}^{3}\ln  \left( b+r \right) +6\,{b}^{2}r_{{0}}\ln 
 \left( b+r \right) +3\,{b}^{2}r\ln  \left( b+r \right) +3\,b{r_{{0}}}
^{2}\ln  \left( b+r \right) 
\nonumber \\
+6\,br_{{0}}r\ln  \left( b+r \right) +3\,{
r_{{0}}}^{2}r\ln  \left( b+r \right) -3\,{b}^{3}\ln  \left( b+r_{{0}}
 \right) -6\,{b}^{2}r_{{0}}\ln  \left( b+r_{{0}} \right) -3\,{b}^{2}r
\ln  \left( b+r_{{0}} \right) 
\nonumber \\
-3\,b{r_{{0}}}^{2}\ln  \left( b+r_{{0}}
 \right) -6\,br_{{0}}r\ln  \left( b+r_{{0}} \right)
 -3\,{r_{{0}}}^{2}r
\ln  \left( b+r_{{0}} \right) +3\,r_{{0}}{b}^{2}-3\,{b}^{2}r
\nonumber  \\
+4\,{r_{{0
}}}^{2}b-3\,br_{{0}}r+{r_{{0}}}^{2}r
\Big )^{1/2}
\quad .
\end{eqnarray}
This differential equation
does not have an analytical solution,
so we  
present the approximation arising from the fourth order Taylor series:
\begin{equation}
r(t;r_0,v_0,t_0,b) \approx
r_{{0}}+v_{{0}} \left( t-t_{{0}} \right) -\frac{3}{4}\,{\frac {{v_{{0}}}^{2}
 \left( t-t_{{0}} \right) ^{2}}{r_{{0}}}}+\frac{1}{4}\,{\frac {{v_{{0}}}^{3}
 \left( 5\,b+7\,r_{{0}} \right)  \left( t-t_{{0}} \right) ^{3}}{{r_{{0
}}}^{2} \left( b+r_{{0}} \right) }}
\quad  .
\label{rttaylornfw}
\end{equation}

\section{Astrophysical applications}

\label{section_astro}

We now test the reliability of the numerical and approximate 
solutions on four SNRs: Tycho, 
see \cite{Williams2016},
Cas A, see \cite{Patnaude2009},  
Cygnus loop,  see \cite{Chiad2015},
and  SN~1006, see \cite{Uchida2013}.
The three astronomically measurable parameters 
are the time since the explosion in years, $t$,
the actual observed radius in pc, $r$,
and the present velocity of expansion in 
km\,s$^{-1}$, see Table \ref{tablesnrs}.
\begin{table}[ht!]
\caption {
Observed astronomical parameters of the SNRs
}
\label{tablesnrs}
\begin{center}
\begin{tabular}{|c|c|c|c|c|}
\hline
Name  & Age (yr)   &  Radius (pc) & Velocity (km\,s$^{-1}$)& References\\
\hline                                                   
Tycho        & 442        &  3.7         & 5300 & Williams~et~al.~(2016)    \\
Cas ~A       & 328        &  2.5         & 4700 & Patnaude~and~Fesen~(2009) \\
Cygnus~loop  & 17000      &  24.25       & 250  & Chiad~et~al.~(2015)       \\
SN ~1006     & 1000       &  10.19       & 3100 & Uchida~et~al.~(2013)       \\
\hline
\end{tabular}
\end{center}
\end{table}
The astrophysical  units   are 
pc for length  and  yr for time.
With these units, the initial velocity is 
$v_0(km s^{-1})= 9.7968 \, 10^5 v_0(pc\,yr^{-1})$.
In all the models here considered, the initial velocity, $v_0$, is constant
in the time interval $[0-t_0]$.

The    goodness of the model  is evaluated
through the percentage  error $\delta_r$
of the radius,
 which is
\begin{equation}
\delta_r = \frac{\big | r_{theo} - r_{obs} \big |}
{r_{obs}} \times 100
\quad ,
\end{equation}
where ${r_{obs}}$ is the radius of the SNR as given by the 
astronomical observations and 
${r_{theo}}$ is the radius suggested by the
model.
In an analogous way, we can define the percentage error  of the velocity. 
Another useful astrophysical variable  is the predicted decrease in 
the theoretical velocity in 10 years,
$\Delta_{10}\,v (km\,s^{-1}) $.

\subsection{Constant density}

The numerical results   for the medium with constant density 
are presented  in Table \ref{tableconstant}.
\begin{table}[ht!]
\caption {
Theoretical  parameters of the SNRs
for the  equation of motion in the case 
of conservation of energy with constant density,
see Section \ref{section_constant}. 
}
k\label{tableconstant}
\begin{center}
\begin{tabular}{|c|c|c|c|c|c|c|}
\hline
Name         &$t_0$ (yr)&$r_0$ (pc)&$v_0 (km\,s^{-1})$& $\delta_r\,(\%)$
& $\delta_v\,(\%)$
& $\Delta_{10}\,v (km\,s^{-1}) $ \\
\hline                                                   
Tycho        & 28.41 &  0.87     & 30000   &   0.1   &  35.55 &  -47.33   \\
Cas ~A       & 17.96    &  0.55  & 30000   &   0.095 & 34.22 & -57.03   \\
Cygnus~loop  & 55.51    &  1.7   & 30000   &    0.23 &  123.5& -0.197  \\
SN ~1006     & 91.43    & 2.79   & 30000   &     0.8 & 37.52 & -26.83  \\
\hline
\end{tabular}
\end{center}
\end{table}

\subsection{Power law densities}

The results for a 
medium with an hyperbolic  density
are presented in  Table \ref{table_hyperbolic},   
\begin{table}[ht!]
\caption {
Theoretical  parameters of the SNRs
for the  equation of motion in the case 
of conservation of energy with an hyperbolic profile of density,
see Section \ref{section_hyperbolic}. 
}
\label{table_hyperbolic}
\begin{center}
\begin{tabular}{|c|c|c|c|c|c|c|}
\hline
Name         &$t_0$ (yr)&$r_0$ (pc)&$v_0 (km\,s^{-1})$& $\delta_r\,(\%)$
& $\delta_v\,(\%)$
& $\Delta_{10}\,v (km\,s^{-1}) $ \\
\hline                                                   
Tycho        & 20.24    &  0.62  & 30000   &   0.017   &  22.2    & -46.53   \\
Cas ~A       & 12.40    &  0.38  & 30000   &   0.127   &  20.37   & -56.4   \\
Cygnus~loop  & 22.85    &  0.7   & 30000   &   0.61    &  181     &  -0.2  \\
SN ~1006     & 68.57    &   2.09 & 30000   &   0.27    &  63.38   & -25.76  \\
\hline
\end{tabular}
\end{center}
\end{table}
those 
for the medium with an inverse square profile of density 
are presented  in Table \ref{table_invsquare},   
\begin{table}[ht!]
\caption {
Theoretical  parameters of the SNRs
for the  equation of motion in the case 
of conservation of energy with an inverse square  profile of density,
see Section \ref{section_invsquare}. 
}
\label{table_invsquare}
\begin{center}
\begin{tabular}{|c|c|c|c|c|c|c|}
\hline
Name         &$t_0$ (yr)&$r_0$ (pc)&$v_0(km\,s^{-1})$& $\delta_r\,(\%)$
& $\delta_v\,(\%)$
& $\Delta_{10}\,v (km\,s^{-1}) $ \\
\hline                                                   
Tycho        & 10.44    &  0.32  & 30000   &   0.016   &  0.98    & -39.7   \\
Cas ~A       & 6        &  0.184 & 30000   &   0.216   &  2.40    & -48.62  \\
Cygnus~loop  &2.28      &  0.07  & 30000   &    0.1    &  272     & -0.18  \\
SN ~1006     & 40.82    &  1.25  & 30000   &   0.089   &  104     & -21.6  \\
\hline
\end{tabular}
\end{center}
\end{table}
and those 
for the medium with an inverse power law  profile of density 
are presented in Table \ref{table_powerlaw}.   
\begin{table}[ht!]
\caption {
Theoretical  parameters of the SNRs
for the  equation of motion in the case 
of conservation of energy with a power law   
profile of density when $\alpha=1.5$,
see Section \ref{section_powerlaw}. 
}
\label{table_powerlaw}
\begin{center}
\begin{tabular}{|c|c|c|c|c|c|c|}
\hline
Name         &$t_0$ (yr)&$r_0$ (pc)&$v_0(km\,s^{-1})$& $\delta_r\,(\%)$
& $\delta_v\,(\%)$
& $\Delta_{10}\,v (km\,s^{-1}) $ \\
\hline                                                   
Tycho        & 15.6     &  0.47  & 30000   &   0.152   &  12.83   & -44.41   \\
Cas ~A       & 9.3      &  0.285 & 30000   &   0.0383  &  40.43   & -47.15   \\
Cygnus~loop  & 9.96     &  0.3   & 30000   &   0.0443  &  23.29   & -0.1     \\
SN ~1006     & 55.15    &  1.689 & 30000   &   0.07    &  31.53   & -22.91 \\
\hline
\end{tabular}
\end{center}
\end{table}

\begin{figure*}
\begin{center}
\includegraphics[width=5cm,angle=-90]{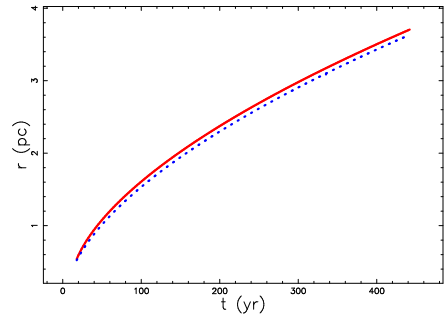}
\end {center}
\caption
{
Numerical solution (full red line) and  asymptotic 
approximate solution 
(blue dashed  line)
for the inverse power law with $\alpha=1.5$
Parameters  as  in Table \ref{table_powerlaw} for Tycho.
}
\label{alpha_asympt}
    \end{figure*}

\begin{figure*}
\begin{center}
\includegraphics[width=5cm,angle=-90]{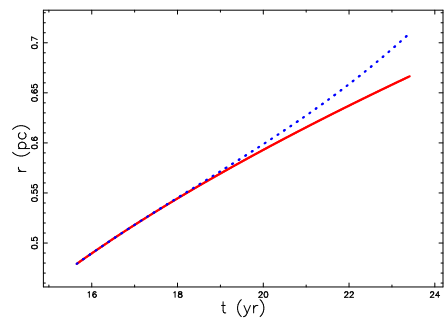}
\end {center}
\caption
{
Numerical solution (full red line) and  Taylor  approximation 
(blue dashed line)
for the inverse power law with $\alpha=1.5$.
Parameters  as  in Table \ref{table_powerlaw} for Tycho.
}
\label{alpha_taylor}
    \end{figure*}

In the case of a density which decreases with a power law profile 
we have already pointed out the absence of an analytical  
solution.
As a consequence, Figure \ref{alpha_asympt} presents
the asymptotic approximate trajectory
as  given by (\ref{rtalphaasympt})
for Tycho  in the  full range of time  $[15.6\,yr-442\,yr]$.
Figure \ref{alpha_taylor} presents
the Taylor approximation of the trajectory 
as  given by (\ref{rttayloralpha})
in the
restricted range of time  $[15.6\,yr-24\,yr]$.

\subsection{Presence of an exponential}

\begin{figure*}
\begin{center}
\includegraphics[width=5cm,angle=-90]{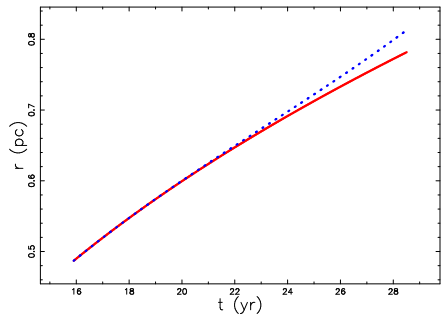}
\end {center}
\caption
{
Numerical solution (full red line) and  Taylor  approximation 
(blue dashed line)
for the exponential profile.
Parameters  as  in Table \ref{table_exponential} for Tycho.
}
\label{exp_taylor}
    \end{figure*}

The astrophysical parameters for an exponential 
profile of density are presented in Table \ref{table_exponential}
and the fit of the trajectory  with
a Taylor expansion, see Equation (\ref{rttaylorexp}),
is presented in Figure \ref{exp_taylor}.

\begin{table}[ht!]
\caption {
Theoretical  parameters of the SNRs
for the  equation of motion in the case 
of conservation of energy with an 
exponential   
profile of density,
see Section \ref{section_exponential}. 
}
\label{table_exponential}
\begin{center}
\begin{tabular}{|c|c|c|c|c|c|c|c|}
\hline
Name         &$t_0$ (yr)&$r_0$ (pc)&b& $v_0(km\,s^{-1})$& $\delta_r\,(\%)$
& $\delta_v\,(\%)$
& $\Delta_{10}\,v (km\,s^{-1}) $ \\
\hline                                                   
Tycho       & 15.83 &  0.48 & 1  & 30000   & 0.22   &  8.12    & -27.62   \\
Cas ~A      & 11.91 &  0.365& 1  & 30000   & 0.29   &  15.27   &-43.88  \\
Cygnus~loop & 5.15  &  0.15 & 0.7& 30000   & 0.085  &  425     & 0   \\
SN ~1006    & 18.35 &  0.56 & 0.7& 30000   & 0.46   &  178     & -0.02  \\
\hline
\end{tabular}
\end{center}
\end{table}

\begin{figure*}
\begin{center}
\includegraphics[width=5cm,angle=-90]{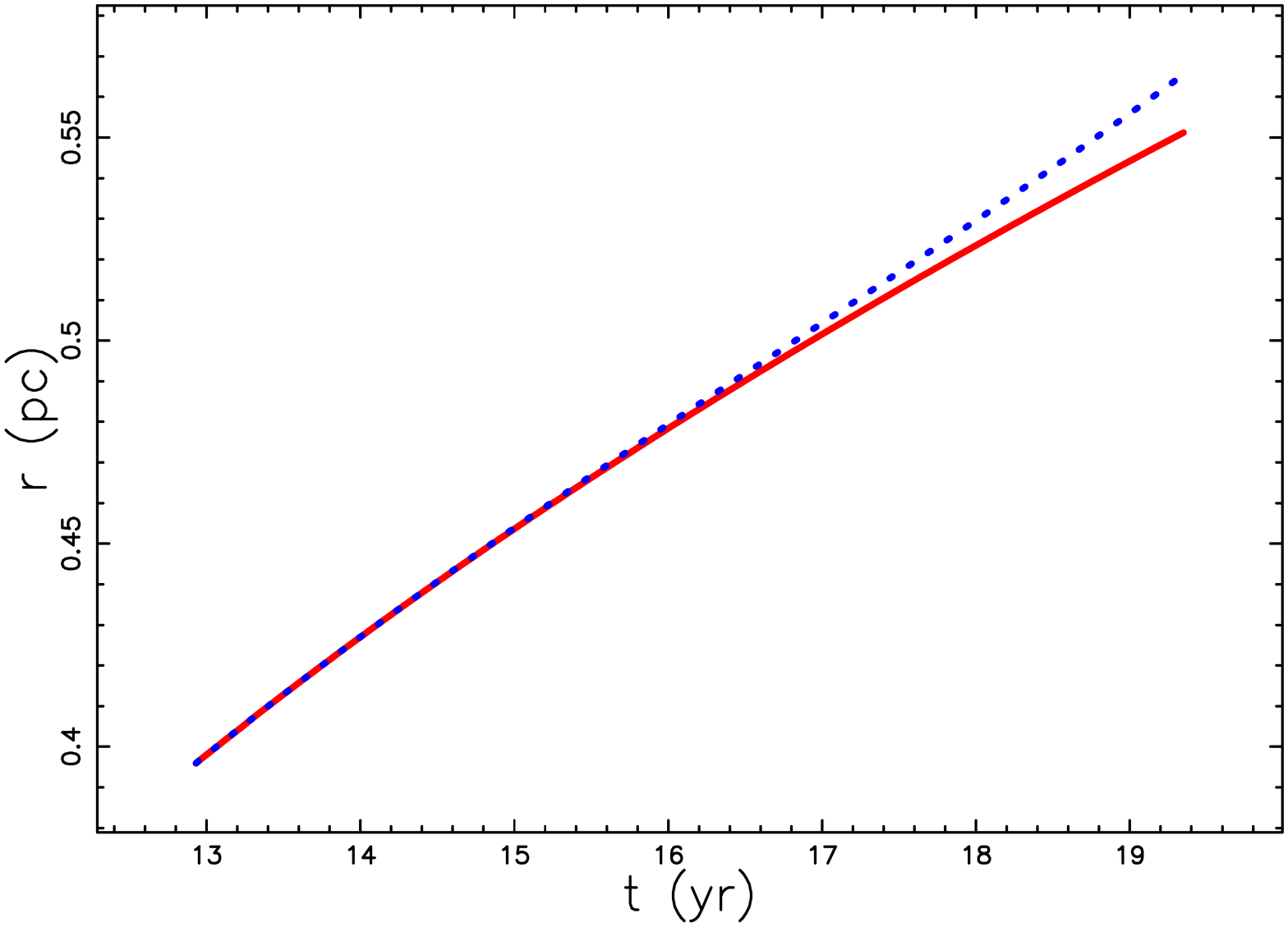}
\end {center}
\caption
{
Numerical solution (full red line) and  Taylor  approximation 
(blue dashed line)
for the Gaussian  profile.
Parameters  as  in Table \ref{table_gaussian} for Tycho.
}
\label{gauss_taylor}
    \end{figure*}

The astrophysical parameters for a Gaussian  
profile of density are presented in Table \ref{table_gaussian}
and the fit of the trajectory  with
a Taylor expansion, see Equation (\ref{rttaylorgauss}),
is presented in Figure \ref{gauss_taylor}.

\begin{table}[ht!]
\caption {
Theoretical  parameters of the SNRs
for the  equation of motion in the case 
of conservation of energy with a 
Gaussian 
profile of density,
see Section \ref{section_gaussian}. 
}
\label{table_gaussian}
\begin{center}
\begin{tabular}{|c|c|c|c|c|c|c|c|}
\hline
Name         &$t_0$ (yr)&$r_0$ (pc)&b& $v_0(km\,s^{-1})$& $\delta_r\,(\%)$
& $\delta_v\,(\%)$
& $\Delta_{10}\,v (km\,s^{-1}) $ \\
\hline                                                   
Tycho       & 12.89 &  0.395 & 1  & 30000   &  0.013  &   21.62   & -0.005   \\
Cas ~A      & 10.95 &  0.335 & 1  & 30000   & 0.034   &   7.79    & -3.2  \\
Cygnus~loop & 3.2   &  0.0979& 0.7& 30000   & 0.0385  &   445     &   0  \\
SN ~1006    & 11.73 &  0.359 & 0.7& 30000   &  0.087  &  206.2    & 0  \\
\hline
\end{tabular}
\end{center}
\end{table}

\subsection{Autogravitating medium}

\begin{figure*}
\begin{center}
\includegraphics[width=5cm,angle=-90]{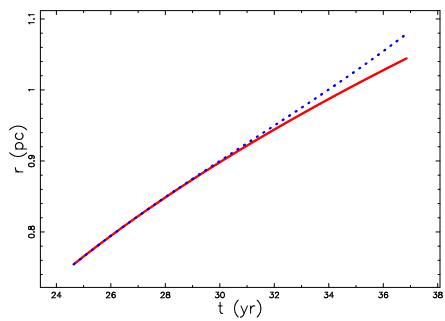}
\end {center}
\caption
{
Numerical solution (full red line) and  Taylor  approximation 
(blue dashed line)
for the autogravitating   profile.
Parameters  as  in Table \ref{table_sech2} for Tycho.
}
\label{sech2_taylor}
    \end{figure*}

The astrophysical parameters for an autogravitating medium  
are presented in Table \ref{table_sech2}
and the fit of the trajectory  with
a Taylor expansion, see Equation (\ref{rttaylorsech2}),
is presented in Figure \ref{sech2_taylor}.

\begin{table}[ht!]
\caption {
Theoretical  parameters of the SNRs
for the  equation of motion in the case 
of conservation of energy with an
autogravitating 
profile of density,
see Section \ref{section_sech2}. 
}
\label{table_sech2}
\begin{center}
\begin{tabular}{|c|c|c|c|c|c|c|c|}
\hline
Name         &$t_0$ (yr)&$r_0$ (pc)&b& $v_0(km\,s^{-1})$& $\delta_r\,(\%)$
& $\delta_v\,(\%)$
& $\Delta_{10}\,v (km\,s^{-1}) $ \\
\hline                                                   
Tycho       & 24.57&  0.752 &   1.5 &   30000 & 0.019 & 25.1 &  -38.3 \\
Cas ~A      & 15.4 &  0.474 &   1   &   30000 & 0.03  & 23.3 &  -45.9 \\
Cygnus~loop & 10.6 &  0.326 &   1   &   30000 & 0.046 & 403  &  -0.03 \\
SN ~1006    & 26.8 &  0.82  &   0.7 &   30000 & 0.002 & 174  &  -0.149 \\
\hline
\end{tabular}
\end{center}
\end{table}

\subsection{NFW profile}

\begin{figure*}
\begin{center}
\includegraphics[width=5cm,angle=-90]{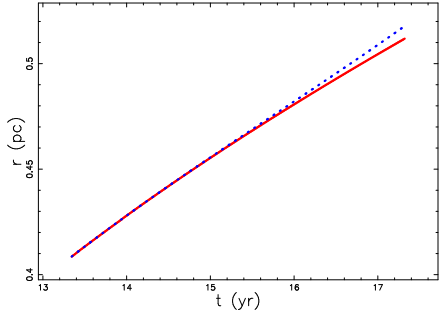}
\end {center}
\caption
{
Numerical solution (full red line) and  Taylor  approximation 
(blue dashed line)
for an NFW    profile.
Parameters  as  in Table \ref{table_nfw} for Tycho.
}
\label{nfw_taylor}
    \end{figure*}

The astrophysical parameters for an  NFW profile of
density   
are presented in Table \ref{table_nfw}
and the fit of the trajectory  with
a Taylor expansion, see Equation (\ref{rttaylornfw}),
is presented in Figure \ref{nfw_taylor}.
\begin{table}[ht!]
\caption {
Theoretical  parameters of the SNRs
for the  equation of motion in the case 
of conservation of energy with an
NFW 
profile of density,
see Section \ref{section_nfw}. 
}
\label{table_nfw}
\begin{center}
\begin{tabular}{|c|c|c|c|c|c|c|c|}
\hline
Name         &$t_0$ (yr)&$r_0$ (pc)&b& $v_0(km\,s^{-1})$& $\delta_r\,(\%)$
& $\delta_v\,(\%)$
& $\Delta_{10}\,v (km\,s^{-1}) $ \\
\hline                                                   
Tycho       & 13.3&   0.408 & 1.5 & 30000 & 0.07  & 3    &  -34.8 \\
Cas ~A      & 8   &   0.245 & 1   & 30000 & 0.073 & 0.26 &  -42.3 \\
Cygnus~loop & 3.43&   0.1052& 1   & 30000 & 0.09  & 338  &  -0.1  \\
SN ~1006    & 27.5&   0.845 & 0.7 & 30000 & 0.074 & 136  &  -14.1 \\
\hline
\end{tabular}
\end{center}
\end{table}

\section{Conclusions}

The thin layer approximation in the framework 
of the conservation of energy is an alternative
to the use of the conservation of momentum in order to find 
the equation of motion for  a supernova remnant (SNR).
In the case where the interstellar medium (ISM) has a  constant density, it is possible
to find the trajectory  in an analytical form, 
see Equation (\ref{energy_rtconstant}).
The case of energy conservation in a medium with 
variable density was also explored but an analytical
trajectory was found only in the case of a medium
characterized by an inverse square decrease of density,
see Equation (\ref{rtinversesquare}).  
The other profiles  of density require a numerical 
integration in order to find the trajectory.
A Taylor series  can provide the  trajectory for a short 
interval of time: see 
Figure  \ref{alpha_taylor} for a power law,
Figure  \ref{exp_taylor}   for an exponential law,
Figure  \ref{gauss_taylor} for  a Gaussian law,
Figure  \ref{sech2_taylor} for  an autogravitating medium 
and
Figure  \ref{nfw_taylor} for  a Navarro--Frenk--White (NFW) density profile.
As an astrophysical target we have chosen to reproduce 
4 standard SNRs.
The match between the observed and simulated radius 
as well as that between  the observed velocity 
and the simulated velocity 
has been 
analysed in  terms of the percentage error,
see Tables 
\ref{tableconstant},
\ref{table_hyperbolic},
\ref{table_invsquare},
\ref{table_powerlaw},
\ref{table_exponential},
\ref{table_gaussian},
\ref{table_sech2} and
\ref{table_nfw}.
Table \ref{synoptic} presents in column
2 the best model for the SNRs here analysed.
\begin{table}[ht!]
\caption {
Synoptical   parameters of the best model
for SNRs with different 
density profiles.
}
\label{synoptic}
\begin{center}
\begin{tabular}{|c|c|c|c|c|c|c|}
\hline
Name  &model       &$t_0$ (yr)&$r_0$ (pc) &$v_0(km\,s^{-1})$& $\delta_r\,(\%)$
& $\delta_v\,(\%)$ \\
\hline                                                   
Tycho       &inverse ~square   & 10.44    &  0.32  & 30000   &   0.016   &  0.98   \\
Cas ~A      &NFW\,,\,b=1\,pc  & 8  & 0.245    & 30000 & 0.073 & 0.26 \\
Cygnus~loop &power law & 9.96  & 0.3 &   30000   &   0.0443  &  23.29   \\
SN ~1006     &power law& 55.15 & 1.689 &  30000   &   0.07    &  31.53    \\
\hline
\end{tabular}
\end{center}
\end{table}
The  solution for the velocity to first order 
allows the  insertion of the back reaction, i.e.
the radiative losses, in the equation for the    energy
conservation, 
see equation (\ref{eqnenergyback}),
and as a consequence  the  velocity corrected to second order,
see equation (\ref{vcorrected}).
The radiative losses allow evaluating the length  at which 
the advancing velocity of the SNR is zero.
\providecommand{\newblock}{}

\end{document}